\documentclass{PoS}

\usepackage{graphicx}
\usepackage{txfonts}
\usepackage{amsfonts}
\usepackage{subfigure}
\usepackage{amssymb,amsfonts}

\title{Perspective for optical high-angular resolution follow-up studies of X-raying AGNs}

\ShortTitle{High-angular resolution studies of AGN in the optical}

\author{\speaker{Lucas Labadie}\thanks{}\\
		I. Physikalisches Institut, Universit\"at zu K\"oln, Z\"ulpicher Str. 77, 50937 K\"oln, Germany\\
		E-mail: \email{labadie@ph1.uni-koeln.de}}

\author{Jens Zuther\\
        I. Physikalisches Institut, Universit\"at zu K\"oln, Z\"ulpicher Str. 77, 50937 K\"oln, Germany\\
		}

\author{Andreas Eckart\\
        I. Physikalisches Institut, Universit\"at zu K\"oln, Z\"ulpicher Str. 77, 50937 K\"oln, Germany\\
		Max-Planck-Institut f\"ur Radioastronomie, Auf dem H\"ugel 69, 53121 Bonn, Germany\\
		}

\author{Tim Staley \\
 		Oxford Astrophysics, Denys Wilkinson Building, Keble Road, Oxford OX1 3RH\\
				}

\author{Craig Mackay \\
		Institute of Astronomy, University of Cambridge, Madingley Road, Cambridge CB3 0HA, UK\\
		}

\author{Rafael Rebolo \\
		Instituto de Astrofisica de Canarias, C/ Via Lactea s/n, La Laguna, Tenerife E-38205, Spain\\
		}

\abstract{We explore the scientific potential of next-generation high-angular resolution optical imager to study the AGN/Host connection. The availability of a significant number of X-raying AGN with natural guide stars, allowing for adaptive optics at optical wavelengths, offers an interesting perspective to complement high-resolution work currently done in the near-infrared.
}
\FullConference{
Nuclei of Seyfert galaxies and QSOs - Central engine \& conditions of star formation \\
November 6-8, 2012\\
Max-Planck-Insitut f\"ur Radioastronomie (MPIfR), Bonn, Germany}

\begin{document}

\section{Motivations}

\noindent Observing the heart of astrophysical objects requires high angular resolution offered by large aperture telescopes. In the context of the starburst/AGN connection, a better understanding of the feedback process can be reached by observing the close environment of the nucleus at the diffraction limit of the instrument. From the ground, the limitations imposed by turbulent atmosphere can be overcome by the use of adaptive optics (AO) systems, which provide a real-time restoration of the otherwise corrupted image. 
Initially, AO systems have been developed in the near-IR range (VLT, Keck, LBT, Gemini) where both the existing technology and the dependence of the Fried and coherence time parameters r$_{\rm 0}$, $\tau_{\rm 0}$ favored longer wavelengths ($\propto$\,$\lambda^{6/5}$). \\
However, the strong improvement in optical sensor sensitivity, quantum efficiency and read-out noise reduction in the last years has profited to the re-emergence of speckle imaging techniques, which -- in conjunction with low-order correcting AO systems -- offer the potential for multi-wavelengths high-angular resolution imaging across the visible to near-infrared range, bringing new insight into the AGN/Host Galaxy connection. 

\section{Methodology}

\begin{figure}[b]
\centering
\includegraphics[width=7.5cm]{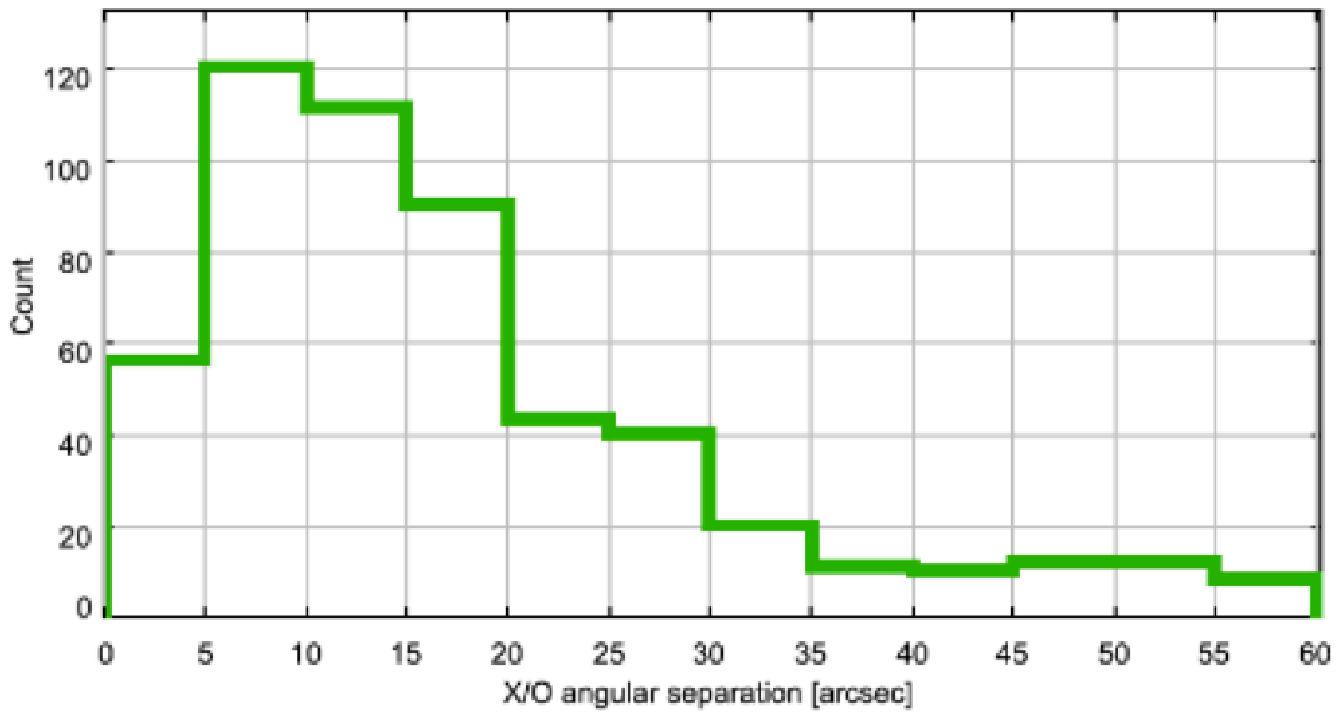}
\hspace{0.5cm}
\includegraphics[width=6.5cm]{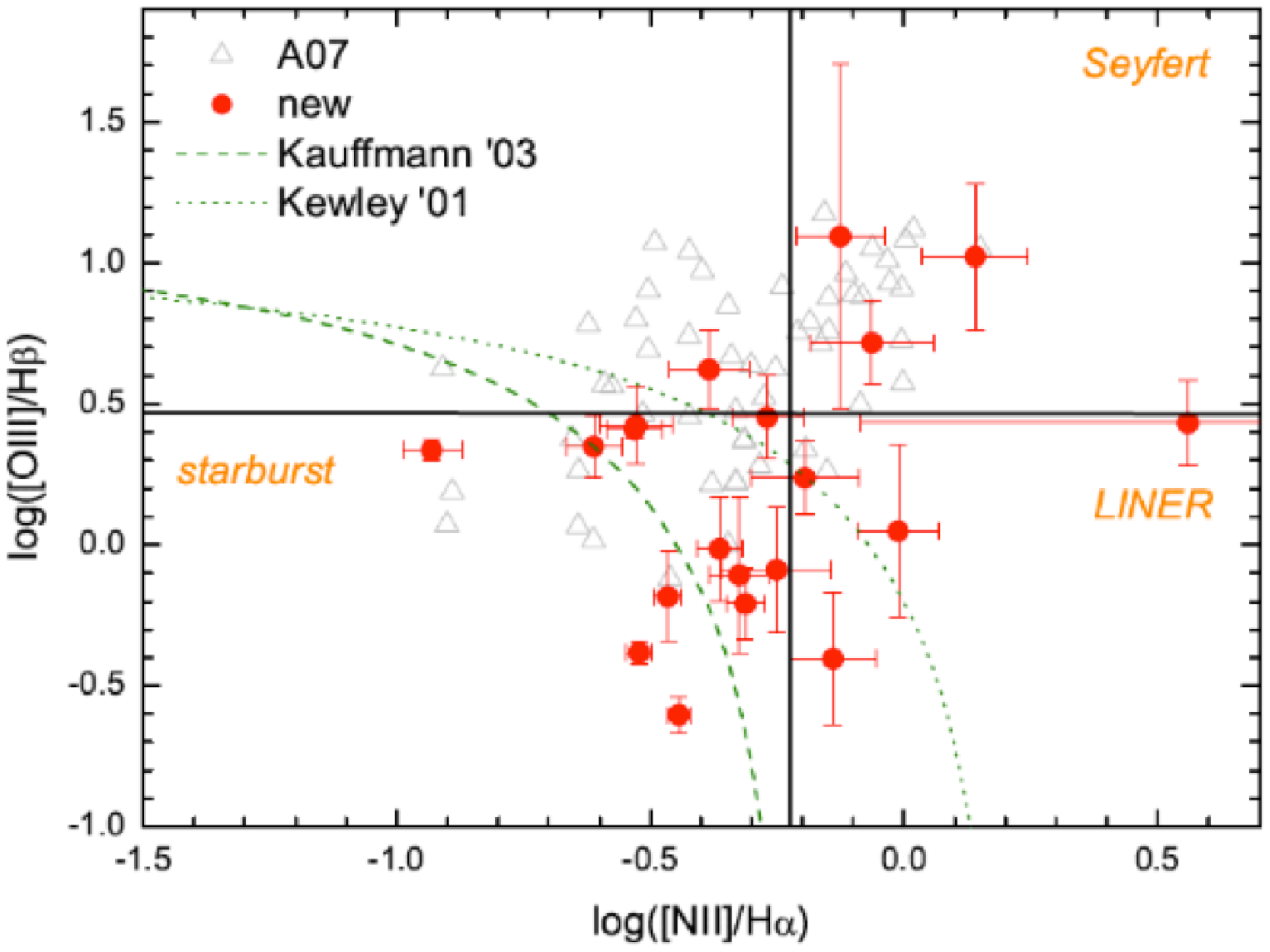}
\caption{
{\it Left:} Angular separation of optical counterparts from the X-ray source. Objects within 20$^{\prime\prime}$ can be considered good matches.
{\it Right:} Optical pectroscopic diagnostics (BPT diagram) of matched, AO-suitable targets. Note the composite region framed by the green curves.}\label{Ima1}
\end{figure}

\noindent Natural Guide Star (NGS) AO is still the reliable working horse at most observatories. What is needed is to find pairs of science targets and nearby guide stars within the isoplanatic patch. Bright enough science targets can certainly be used for self-referencing.
We have used Virtual Observatory (VO) techniques to prepare a sample of NGS-AO-suitable X-ray bright AGN. The cross match is based on the SDSS, ROSAT, and FIRST public databases. We carried out correlation analyses to achieve a reliable classification of sources, obtaining a sample of circa 500 objects. The resulting detailed sample can then be used for high-angular- resolution follow-up studies with narrowband and broadband optical imaging. The left panel of Fig.~\ref{Ima1} points the high number of available source/AO-star couples within an acceptable isoplanatic patch angle. The right panel of Fig.~\ref{Ima1} displays the current classification in the BPT diagram of the sample members based on seeing-limited optical spectroscopic studies.

\section{Examples}

\begin{figure}[h]
\centering
\includegraphics[width=13.25cm]{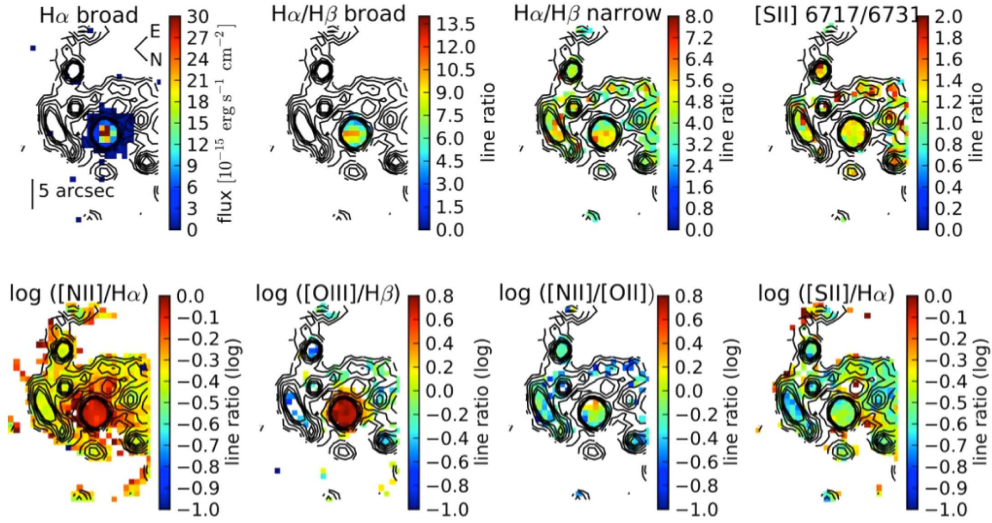}
\caption{Maps of the broad H$\alpha$ component, the flux ratios of the broad and narrow H$\alpha$ and H$\beta$ components, the flux ratio of [SII]$\lambda$6717\,/\,[SII]$\lambda$6731 and the logarithm of the flux ratios of [NII]\,$\lambda$6583\,/\,H$\alpha$, [OIII]$\lambda$5007\,/\,H$\beta$, [NII]$\lambda$6583\,/\,[OII]$\lambda$3727,29, and [SII]$\lambda$6717,6731/H$\alpha$. (from Scharw\"achter et al.\,\cite{Scharw2011}).
}\label{Ima2}
\end{figure}

\noindent 
Broadband and narrow band {\it optical} imaging coupled to AO high-angular resolution techniques allows to resolve the nuclear structure of Active Galaxies and reveal the morphology of their hosts. Where appropriate filters are available for [OIII], [NII], [SII] or H$\alpha$ at a galaxy's redshift, AGN-ionized clouds yet unknown or targeted by SDSS can be revealed in great details (cf. Keel\,et\,al.\,\cite{Keel2011}). Diffraction-limited optical imaging on a 4.2-m class telescope delivers a 40 mas resolution at 0.6\,$\mu$m, which permits to probe such structures at 30\,pc scales at z=0.04. Galaxies located in different regions of the BPT diagram can be imaged under the form of the [OIII]/H$\beta$ and [NII]/H$\alpha$ line diagnostics and therefore resolve the AGN, composite and HII extended narrow-line region (cf. Fig.~\ref{Ima2}\,\cite{Scharw2011}).

\section{Instrumentation}

\begin{figure}[h]
\centering
\includegraphics[width=10.5cm]{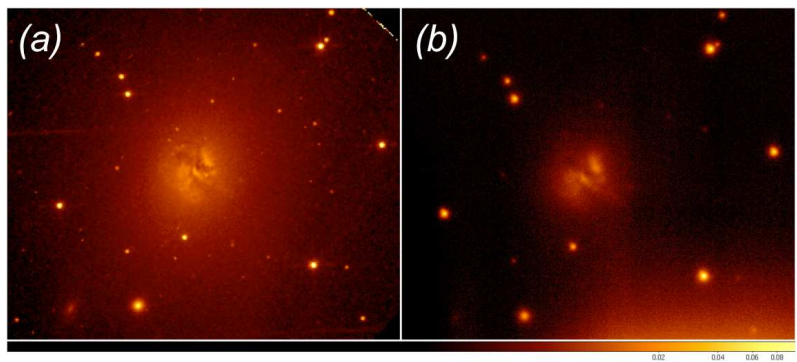}
\caption{Comparison of optical images of 3C 405 (Cyg~A) taken in the V band (0.55 mm) with HST/ACS (a) and in the I band (0.79 mm) with the 2.5-m NOT telescope (b). Because it is recorded at a shorter wavelength, the resolution of the HST image is slightly better.}\label{Ima3}
\end{figure}

\noindent Ground-based diffraction-limited imaging at optical wavelengths has been achieved in a recent past through speckle techniques, but was restricted to bright sources due to limited read-out performances of the detectors and to the spread of the speckle cloud as the telescope diameter goes significantly larger (typically $\ge$2.5\,m) than the turbulence cells. With the next-to-come deployment of the Adaptive Optics Lucky Imager (AOLI) at the 4.2-m WHT\,\cite{Mackay2012} we plan to combine new generation photon counting EMCCD with low-order curvature wavefront sensors to allow much fainter reference stars, increased isoplanatic patch thanks to selection of ``lucky frames'', and therefore larger sky coverage even at high galactic latitudes. AOLI, planned for operation in late 2013\,/\,early 2014, is built on the heritage of science instruments such as LuckyCam\,\cite{Law2009} and FastCam\,\cite{Labadie2010}.\\ 
AOLI is composed of a broadband AO optical wavefront sensor channel and a science camera composed of four 1024x1024 EMCCDs. Classical Shack-Hartmann-based AO systems like PALMAO split the light into many sub- pupils to sense the wavefront, limiting us to reference stars of magnitude I\,$\sim$\,12--14. In addition, high-order wavefront sensors can access in the visible only small isoplanatic angles of few arcseconds. At the contrary, theoretical studies\,\cite{Racine2006} suggest that curvature wavefront sensors are more sensitive, especially in sensing the low-order errors due to turbulence. The AO system of AOLI consists of a non-linear curvature WFS and a low order AO corrector using a DM coupled to a wide-field array detector. Our simulations suggest we will be able to pick-up a reference star of about I$\sim$17.5, which greatly increases the sky coverage. To allow larger FOV than few arcseconds, the science camera of AOLI is composed of four detectors in a ``buttable setup'' thanks to a pyramid mirror where the magnified image is projected onto. Post-selection of those fast--acquired frames (10--30\,ms) less affected by the turbulence allows to increase the isoplanatic patch to $\sim$60$^{\prime\prime}$. The design of the science camera allows to include different filters in the four different quadrants. Depending on the selected magnification, a FOV of 120$^{\prime\prime}$$\times$120$^{\prime\prime}$ down to 12$^{\prime\prime}$$\times$12$^{\prime\prime}$ can be selected. Fig.~\ref{Ima3} shows how optical images with unprecedented angular resolution can be acquired from the ground, hence opening new doors in the field of Active Galaxies.\newline



\end{document}